\documentclass[11pt,a4paper]{article}
\usepackage[latin1]{inputenc}
\usepackage{jcappub}
\usepackage{amsmath}
\usepackage{amsfonts}
\usepackage{amssymb}
\usepackage{amsthm}
\usepackage{graphicx}
\usepackage{epsfig}
\usepackage{color}
\usepackage{amssymb}
\usepackage{mathtools}

\def\beq{\begin{equation}}
\def\eeq{\end{equation}}
\def\baq{\begin{eqnarray}}
\def\eaq{\end{eqnarray}}
\newcommand{\ee}[1]{\begin{equation}#1\end{equation}}
\newcommand{\ea}[1]{\begin{align}#1\end{align}}

\providecommand{\f}[2]{\frac{{#1}}{{#2}}}

\newcommand{\h}{\alpha}

\title{Do metric fluctuations affect the Higgs dynamics during inflation?}

\author[a]{Tommi Markkanen,}
\author[b]{Sami Nurmi,}
\author[c]{and Arttu Rajantie }

\affiliation[a]{Department of Physics, King's College London, Strand, London WC2R 2LS, UK}
\affiliation[b]{Department of Physics, University of Jyv\"{a}skyl\"{a},  P.O. Box 35, FI-40014 University of Jyv\"{a}skyl\"{a}, Finland}
\affiliation[c]{Department of Physics, Imperial College London, SW7 2AZ, United Kingdom}

\abstract{We show that the dynamics of the Higgs field during inflation is not affected by metric fluctuations if the Higgs is an energetically subdominant light spectator. For Standard Model parameters we find that couplings between Higgs and metric fluctuations are suppressed by $\mathcal{O}(10^{-7})$. They are negligible compared to both pure Higgs terms in the effective potential and the unavoidable non-minimal Higgs coupling to background scalar curvature. The question of the electroweak vacuum instability during high energy scale inflation can therefore be studied consistently using the Jordan frame action in a Friedmann--Lema\^{i}tre--Robertson--Walker metric, where the Higgs-curvature coupling enters as an effective mass contribution.
Similar results apply for other light spectator scalar fields during inflation.}

\emailAdd{tommi.markkanen@kcl.ac.uk}
\emailAdd{sami.t.nurmi@jyu.fi}
\emailAdd{a.rajantie@imperial.ac.uk}

\begin{document}
\begin{flushleft}
	\hfill		 KCL-PH-TH/2017-32
	\newline \hfill	\raggedleft		 IMPERIAL/TP/2017/TM/02
\end{flushleft}
\maketitle

\section{Introduction}

A striking feature of the Standard Model (SM) of particle physics is that current observations favour a metastable electroweak (EW) vacuum which should decay given sufficiently long time \cite{Bednyakov:2015sca,Degrassi:2012ry,Buttazzo:2013uya,Bezrukov:2012sa,Espinosa:1995se, Isidori:2001bm,Ellis:2009tp}. Its decay rate today is negligible but during inflation the EW vacuum could easily have been destabilised in less than a Hubble time due to fluctuations in the Higgs field. As an experimental observation we know that the EW vacuum must survive the extreme conditions of the Early Universe providing a cosmological constraint for the SM, and any theory beyond, which is currently an actively studied topic \cite{vacstab,vacstab1,vacstab2,vacstab3,vacstab4,vacstab5,vacstab6,vacstab7,vacstab8,vacstab9, vacstab10,vacstab11,vacstab12,vacstab13,vacstab14,vacstab15,vacstab16,vacstab17,vacstab18, vacstab19,vacstab20,vacstab21,Moss:2015gua,Figueroa:2017slm}. 
In Ref. \cite{Herranen:2015ima} the reheating epoch was shown also to be capable of triggering the instability, which has subsequently been investigated in \cite{Postma:2017hbkf,Ema:2017loe,Ema:2016kpf,Kohri:2016wof,Enqvist:2016mqj,Ema:2017rkk}. The instability has furthermore recently been shown to be enhanced by the presence black holes as they can act as nucleation sites for vacuum decay \cite{Gregory:2013hja,Burda:2015isa,Burda:2015yfa,Burda:2016mou,Mukaida}.

Due to the unavoidable coupling of the Higgs field to the scalar curvature of gravity the SM alone can in fact be stable during and after inflation \cite{vacstab7,Herranen:2015ima}. The stability/instability very much depends on the precise value of the non-minimal coupling to curvature $\xi$ which in curved space is always generated by renormalization group running induced by the changing background curvature \cite{vacstab7}. Hence any new physics stabilising the EW vacuum must take into account the modification to the running of $\xi$ and its potential significance to vacuum stability.

In the vast majority of works studying vacuum stabilty in the early Universe it is simply assumed that fluctuations of the metric are insignificant and it is only necessary to quantize the matter degrees of freedom paralleling the assumptions usually also made when analysing the dynamics of a spectator fields in general \cite{Enqvist:2013qba,Enqvist:2013kaa,Nurmi:2015ema, Enqvist:2015sua,Kainulainen:2016vzv,Heikinheimo:2016yds,Hardwick:2017fjo,Hardwick:2017qcw}. In this approach, the non-minimal coupling $\xi$ enters as an effective mass term. However, this was challenged recently in Ref.  \cite{Calmet:2017hja}, which argued that it is incomplete and that the metric fluctuations have to be included.

In this work we address the issue of significance of metric fluctuations for the specific case of a subdomimant spectator Higgs field. Our investigation is performed first at the background level and finally as a full Arnowitt--Deser--Misner (ADM) fluctuation analysis to quadratic order. Our primary goal is to conclusively determine how good an approximation is treating gravity as a classical background when discussing the potential EW vacuum instability during high scale inflation. A related issue arising from the non-minimal coupling to scalar curvature is the potential frame dependence of the results and choosing the correct frame for quantization, which has garnered considerable debate over the years \cite{Postma:2014vaa,Calmet,Sasaki, Capozziello2,Faraoni,Kamenshchik:2014waa,Pandey:2016unk,Pandey:2016jmv}. Currently there is no apparent consensus whether or not theories quantised in the Einstein or Jordan frame are equivalent, see for example the opposing conclusions of \cite{Postma:2014vaa} and \cite{Kamenshchik:2014waa}. 

Our conclusion is that when the Higgs field is subdominant, the dynamics of its fluctuations do not depend on the choice of the frame or whether the metric fluctuations are included.
\section{The setup}

We revisit the stability analysis of the EW vacuum during inflation relaxing the assumption of a fixed gravitational background in the computation of the effective Higgs potential. We consider a setup where slow-roll inflation is driven by a singlet scalar inflaton field $\phi$ and the rest of the matter fields is described by the SM. We assume the action is given by  
\ea{
\label{S_intro}
S=\int d^4 x \sqrt{-g}\bigg(&\f{R}{2}M_{\rm pl}^2-\f{1}{2}\nabla^{\mu}\phi\nabla_{\mu}\phi -V(\phi)-\f{1}{2}\nabla^{\mu}h\nabla_{\mu} h -U(h) -\f{1}{2}\xi R h^2\nonumber \\&-\frac{1}{2}\lambda_{\phi h}h^2 \phi^2+ {\cal L}_{\rm SM}\bigg)~,
} 	
where $h = \sqrt{2 H^{\dag}H}$ is the norm of the SM Higgs doublet
\beq 
\label{higgspot}
U(h) =-\f{1}{2}m^2 h^2+ \f{1}{4}\lambda h^4\ ,
\eeq
and ${\cal L}_{\rm SM}$ stands for the rest of the SM Lagrangian which includes also all the remaining degrees of freedom of $H$. Generically, the SM Higgs is an energetically subdominant spectator field\footnote{Note that Higgs fluctuations generated during inflation do not drive the field to super-Planckian field values if the electroweak vacuum remains stable. The effective mass from either self-interactions or  the (positive) non-minimal coupling suppresses fluctuations above $H/\lambda_{h}^{1/4}$ or  $H/\xi$, depending on which one is smaller.} during inflation \cite{Enqvist:2013kaa}. In the following we will always assume this is the case and the energy density of universe $\rho_{\rm tot}$ is dominated by the energy density of the inflaton field $\rho_{\rm tot}\simeq \rho_{\phi}\gg \rho_{h}$. 

Note that we have not included a $\xi_{\phi} R \phi^2$ term or any other direct curvature couplings for the inflaton field. A crucial difference compared to spectator fields is the non-renormalisability of the inflaton sector which dominates the energy density.  Metric fluctuations cannot be neglected in computing radiative corrections to the inflaton potential. Consequently, the quantum corrected action for the inflaton plus gravity at high energies is not uniquely determined by low energy data. Here we assume that the full quantum corrected action takes the minimally coupled form of eq. (\ref{S_intro}) at some scale $\mu_{*}$ comparable to $\rho^{1/4}$ during inflation. Curvature couplings can still be generated through radiative corrections at other scales. However, during inflation the inflaton loops  are suppressed by the flatness of its potential (approximative shift symmetry). 
These slow roll suppressed terms can be neglected in our analysis which verifies that the assumption of minimally coupled inflaton sector is self-consistent. 

The role of the non-minimal coupling Higgs coupling is quite different.  Firstly, if the SM fields are quantised as test fields in a classical curved background, the SM sector is renormalisable and one-loop corrections generate the non-minimal coupling $\xi R h^2$. Secondly, its running is not suppressed but radiatively natural values $\xi \sim \beta_{\xi}$ can significantly affect Higgs dynamics during inflation.

It should also be noted that even if there was a non-minimal inflaton coupling $\xi_{\phi} R \phi^2$, it can be removed at any given scale by a conformal transformation. This generates a  $\xi_{\phi}$ dependent non-renormalisable Higgs-inflaton coupling in the Einstein frame. If the coupling is small enough, our analysis is not affected and the results hold true also for the non-minimally coupled inflaton. It would be an interesting topic for a future work to investigate this at a quantitative level.  


The renormalisable Higgs-inflaton coupling  $\lambda_{\phi h}h^2 \phi^2$ is not forbidden by symmetries and should therefore be included according to general principles of effective field theories. However, apart from Planck mass suppressed graviton mediated contributions, its running is proportional to $\lambda_{\phi h}$ only. Therefore, if we impose the renormalization condition $\lambda_{\phi h} (\mu_0)= 0$ at some scale $\mu_0$ the Higgs-inflaton coupling vanishes on all scales (neglecting the graviton contributions)
\beq
\lambda_{\phi h} = 0~.
\eeq
In other words, the choice $\lambda_{\phi h}= 0$ is radiatively stable and in the following we will concentrate on this specific case.  

The non-minimal curvature coupling of the Higgs $\sim\xi R h^2$ can also be viewed as an effective Higgs-inflaton coupling, as we will discuss in more detail below. However, in contrast to $\lambda_{\phi h}=0$, setting $\xi = 0$ is not a radiatively stable choice as was first discovered in \cite{Chernikov:1968zm,Tagirov:1972vv,Callan:1970ze}. While $\xi(\mu)$ can of course be renormalized to zero at any given scale $\mu_0$, radiative corrections from Higgs loops in a curved spacetime drive $\xi(\mu)$ to non-zero values when moving away from $\mu_0$. As shown in \cite{vacstab7} and also discussed in \cite{Herranen:2015ima,vacstab13} the running scale $\mu$ gets a large contribution from the background curvature indicating that a non-zero $\xi$ will always be generated due to the changing Hubble rate of the background, an example of curvature induced running.

At one loop level $\xi = 1/6$ corresponds to the conformal fixed point where the Higgs does not fluctuate at all and the stability of the EW vacuum is trivially maintained during inflation. The same applies to $\xi >1/6$ when the Higgs is effectively massive during inflation and its fluctuations are heavily suppressed. Here we concentrate on the range $0 < \xi < 1/6$ where the fate of the EW vacuum is a non-trivial interplay between the positive contribution to the effective Higgs mass from the non-minimal coupling and negative contribution (on large enough energy scales) from the quartic self-coupling $\lambda$. 

\section{From Jordan to Einstein frame}
\label{sec:JE}
Instead of working in terms of the Jordan frame action  (\ref{S_intro}) used e.g. in \cite{vacstab7} we choose to investigate Higgs fluctuations in the Einstein frame. This is mainly for computational reasons but also serves to explicitly demonstrate the equivalence of the two frames up to non-renormalizable Planck mass suppressed terms.

To switch to the Einstein frame, we perform the standard conformal scaling of the metric  
\beq
\tilde{g}_{\mu\nu}=\Omega^2g_{\mu\nu}~,\qquad \Omega^2=1-\f{\xi h^2}{M_{\rm pl}^2}~,\label{conft}
\eeq
after which the action (\ref{S_intro}) (with $\lambda_{\phi h}=0$) reads\footnote{For the $(+,+,+)$ conventions we are using the conformal transformation (\ref{conft}) gives $\tilde{R}=\Omega^{-2}R-6\Omega^{-3}\Box\Omega$. } 
\ea{
\label{SEinsteinfull}
S=\int d^4 x \sqrt{-\tilde{g}}\bigg[&\f{\tilde{R}}{2}M_{\rm pl}^2-\Omega^{-2} \f{1}{2}\tilde{\nabla}^{\mu}\phi\tilde{\nabla}_{\mu}\phi-\Omega^{-4}V(\phi)\\\nonumber
&-
\Omega^{-2} \f{1}{2}\tilde{\nabla}^{\mu}h\tilde{\nabla}_{\mu}h -\Omega^{-4}U(h)-3\frac{\xi^2h^2}{M_{\rm pl}^2}\Omega^{-4} \tilde{\nabla}^{\mu}h\tilde{\nabla}_{\mu}h~.}
Here $\tilde{R}$ denotes the Einstein frame curvature scalar computed from the rescaled metric $\tilde{g}_{\mu\nu}$ and $\tilde{\nabla}^{\mu}{\phi} = \tilde{g}^{\mu\nu}\partial_{\nu} \phi$. In what follows, all the quantities refer to Einstein frame variable, unless explicitly stated otherwise, and we will simply drop the tildes. 

We concentrate on the limit where Higgs is energetically subdominant, $\rho_{h}\ll \rho_{\phi} \simeq \rho_{\rm tot}$, which implies that $h/M_{\rm pl}\ll 1$. To shorten the notation we denote this ratio by  
\beq
\label{alpha}
\h \equiv \frac{h}{M_{\rm pl}}\ll 1~.
\eeq 
The action (\ref{SEinsteinfull}) can be greatly simplified by expanding in $\h$. To leading order in the expansion we get  
\baq
\label{Sexpand}
S &=&\int {\sqrt{-g}}
\left[\rule{0pt}{4ex}\right.
\frac{1}{2}M_{\rm pl}^2 R - \left(\frac{1}{2}\nabla^{\mu}\phi\nabla_{\mu}\phi+V(\phi)\right) -\left(\frac{1}{2}\nabla^{\mu}h\nabla_{\mu}h+U(h)\right)\\
\nonumber 
&&-\frac{1}{2}\xi\h^2\left(\nabla^{\mu}\phi\nabla_{\mu}\phi+4V(\phi)\right)+{\cal O}\left(\xi\h^2 {\cal L}_{ h}\right)+{\cal O}\left(\xi^2\h^4 {\cal L}_{ \phi}\right)\left]\rule{0pt}{4ex}\right.~,
\eaq
where in the error terms we have denoted ${\cal L}_{\phi} =\frac{1}{2}\nabla^{\mu} \phi\nabla_{\mu}\phi+V(\phi)$ and ${\cal L}_{h} =\frac{1}{2}\nabla^{\mu} h\nabla_{\mu}h+U(h)$. Note that performing the usual canonical normalization for the kinetic term would not have changed the result in (\ref{Sexpand}). 

Although divided by $M_{\rm pl}^2$, the fourth term on the right hand side of (\ref{Sexpand}) is not suppressed compared to ${\cal L}_{h}$ as it is multiplied by a $\phi$ dependent term which during inflation is proportional to $H^2 M_{\rm pl}^2$. Indeed, as we will discuss below, this term is exactly the Jordan frame non-minimal Higgs coupling written in the Einstein frame.

\subsection{Action with the metric and inflaton fluctuations neglected}  

In \cite{vacstab7} and \cite{Herranen:2015ima} the vacuum stability was investigated treating the energetically subdominant Higgs as a test field in a fixed Friedmann--Lema\^{i}tre--Robertson--Walker (FLRW) background, neglecting all metric and inflaton perturbations.  This amounts to using the line element
\beq
ds^2=-dt^2 + a^2(t)d{\bf x}^2\ ,\eeq
and the classical Einstein/Friedmann equations
\beq
3 H^2M^2_{\rm pl}  = \frac{1}{2}\dot{\phi}^2+V(\phi)\ , \qquad 
2\dot{H}M_{\rm pl}^2 = -\dot{\phi}^2\eeq
where $\phi=\phi(t)$, which yields for the scalar curvature
\beq
\label{RforFRW}
R = 12H^2+6\dot{H}= M_{\rm pl}^{-2}(4V(\phi) -\dot{\phi}^2)=M_{\rm pl}^{-2}(\nabla^{\mu}\phi\nabla_{\mu}\phi+4V(\phi))=-M_{\rm pl}^{-2} {T_{\mu}}^\mu\ . 
\eeq
Using these in (\ref{Sexpand}), the Higgs-dependent part of the Einstein frame action becomes 
\beq
\label{S_h}
S_{h} = \int  d^4 x\; a^3(t)\left(\frac{1}{2}\dot{h}^2 - \frac{1}{2}\frac{(\nabla h)^2}{a^2(t)}-U(h)-\frac{\xi}{2}R h^2 +{\cal O}\left(\xi\h^2 {\cal L}_{ h}\right)+{\cal O}\left(\xi^2\h^4{\cal L}_{ \phi}\right)\right)~.
\eeq
Up to the Planck mass suppressed non-renormalisable terms this exactly coincides with the Jordan frame form we started in (\ref{S_intro}) as it should. 

Up to terms $\sim \mathcal{O}(h/M_{\rm pl})$ the equivalence of the two frames when the metric is fixed can easily be demonstrated to be true for all backgrounds and not just ones that are homogeneous and isotropic. This comes by realizing that the last two equalities of (\ref{RforFRW}) do not require the FLRW form of the metric and come via the trace of the Einstein equation and ${G_{\mu}}^\mu=-R$, and furthermore that (\ref{Sexpand}) was also derived without any assumptions of homogeneity or isotropy.

\section{Quadratic action including all fluctuations}

In this section we move beyond the assumption of the metric being a fixed background. As the first step, we need to expand the action (\ref{Sexpand}) to quadratic order around the classical inflationary solutions retaining all scalar perturbations. As the tensor perturbations, i.e. gravitons, do not couple to scalars at this order we can neglect them here \cite{Mukhanov:1990me}. From the quadratic action one may then proceed to compute the one loop renormalisation group improved effective potential for the Higgs field.  

\subsection{Background solutions}

The equations of motion for a classical homogeneous and isotropic background obtained from the action (\ref{Sexpand}) (retaining only the leading part in $\alpha$) are given by 
\baq
\label{phieom}
\left(\ddot{\phi} + 3 H\dot{\phi}\right)(1+\xi \h^2)+2 \xi\h  \frac{\dot{\phi}\dot{h}}{M_{\rm pl}} + \partial_{\phi}V_{\rm tot}&=&0\\
\label{heom} 
\ddot{h} + 3 H\dot{h}+\partial_{h}V_{\rm tot}&=&0\\
\label{friedmann}
\frac{1}{2}\dot{\phi}^2+ V(\phi) +\frac{1}{2}\dot{h}^2+ U(h)+ \xi \h^2 \left(2V(\phi)+\frac{1}{2}\dot{\phi}^2\right) &=& 3 M_{\rm pl}^2 H^2  ~, 
\eaq
where we have defined 
\beq
V_{\rm tot}\equiv V(\phi) + U(h) +\frac{1}{2}\xi\h^2\left(4V(\phi) -\dot{\phi }^2\right)~.
\eeq

We concentrate on inflaton dominated solutions where the Higgs is a dynamically irrelevant spectator, $ \frac{1}{2}\dot{h}^2+U(h)\ll 3H^2M_{\rm pl}^2$. Note that the last term on the left hand side of eq. (\ref{friedmann}) is small due to (\ref{alpha}). We define slow-roll parameters in the usual way 
\beq
\label{srparameters}
\epsilon_{i} = \frac{M_{\rm pl}^2}{2}\left(\frac{\partial_{i}V_{\rm tot}}{3 H^2 M_{\rm pl}^2}\right)^2~,\qquad \eta_{ij} = \frac{\partial_{i}\partial_{j}V_{\rm tot}}{3 H^2}~.
\eeq
For $\epsilon_{\phi}, |\eta_{\phi}|\ll 1$ and $|\eta_h|\ll 1$, the system admits an attractor solution where the dominant inflaton $\phi$ drives standard slow-roll inflation and the subdominant Higgs is a slowly-rolling light spectator field,
\baq
\label{Heq} 
3 H^2 M_{\rm pl}^2 &=& V(\phi)(1+{\cal O}(\epsilon,\rho_h/\rho_{\rm tot}))\\
\label{phieq}
\dot{\phi} &=& -\sqrt{2\epsilon_{\phi}} H M_{\rm pl}\left(1+ {\cal O}(\epsilon,\rho_h/\rho_{\rm tot},\sqrt{\epsilon_h}\xi \h)\right)\\  
\label{chieq}
\dot{h} &=& -\sqrt{2\epsilon_{h}} H M_{\rm pl}(1+{\cal O}(\epsilon))~.   
\eaq
Here ${\cal O}(\epsilon)$ stands for higher order slow-roll corrections, ${\cal O}(\rho_h/\rho_{\rm tot})$ terms, where $\rho_{h} =  \frac{1}{2}\dot{h}^2+U(h)+ \xi\h^2(2 V(\phi) -\dot{\phi}^2/2)$, arise both from the subdominant Higgs contributions in eq. (\ref{friedmann}) and the non-canonical inflaton kinetic term in eq. (\ref{phieom}), and ${\cal O}(\sqrt{\epsilon_h}\xi \h)$ comes from the kinetic term. 

Note that the slow-roll parameter $\epsilon_h$ is generically very small    
\beq
\epsilon_{h} ={\cal O}(\alpha^2 \eta_{h}^2)~. 
\eeq
This follows directly from (\ref{srparameters}) which, using eq. (\ref{RforFRW}) and dropping the negligible mass term (for $h\gg 246$ GeV) from eq. (\ref{higgspot}), gives the expressions 
\begin{align}
\label{epsilon_h}
\sqrt{2 \epsilon_h} &=&\frac{\lambda h^3}{3H^2 M_{\rm pl}}+ 4 \frac{\xi h}{M_{\rm pl}} (1+{\cal O}(\epsilon_{\phi},\rho_h/\rho_{\rm tot}))~,\qquad  \eta_h = \frac{\lambda h^2}{H^2}+ 4 \xi (1+{\cal O}(\epsilon_{\phi},\rho_h/\rho_{\rm tot}))\,.
\end{align}

\subsection{Fluctuations around the classical solution}
\label{sec:fluc}
We now move on to consider small fluctuations around the classical solution $\phi_0,h_0, a_0$ given by eqs. (\ref{Heq}), (\ref{phieq}) and (\ref{chieq}). We expand the inflaton and Higgs fields as 
\beq
\phi(t,x) = \phi_0(t)+\delta\phi(t,x)~,\qquad h(t,x) = h_0(t)+\delta h(t,x)~.
\eeq
For the metric fluctuations, we follow \cite{Maldacena:2002vr} and write the full metric in the ADM form 
\beq
\label{ADM}
ds^2 =-N^2 dt^2 + h_{ij}(dx^i+N^i dt)(dx^j+N^jdt)~,
\eeq
and choose the spatially flat gauge where
\beq
\label{gauge}
h_{ij} = a^2(t)(\delta_{ij}+\gamma_{ij}), \qquad \delta^{ik}\partial_{k}\gamma_{ij}=0,\qquad \delta^{ij}\gamma_{ij}=0~.
\eeq

Substituting the ADM metric (\ref{ADM}) into the action (\ref{Sexpand}) and setting its variation with respect to $N$ and $N^i$ to zero yields four constraint equations. Their solution in the spatially flat gauge is given by 
\beq
\label{constraints}
N = 1 -\sqrt{\frac{\epsilon_{\phi}}{2}}\frac{\delta\phi}{M_{\rm pl}}~,\qquad N^i = \partial^i \psi,\qquad \partial^i\partial_i \psi = \sqrt{\frac{\epsilon_{\phi}}{2}}\frac{\delta\dot{\phi}}{M_{\rm pl}}~.
\eeq
to first order in perturbations and to leading order in slow-roll. To this order, the result coincides with the single-field inflation \cite{Maldacena:2002vr}. Contributions involving the Higgs fluctuations $\delta h$ are proportional to $\sqrt{\epsilon_{h}} = {\cal O}(\alpha \eta_h)$ and therefore slow-roll and $\alpha$ suppressed. The full solution with no slow-roll approximation made is given explicitly in the Appendix \ref{fullaction}. 

Substituting the solution (\ref{constraints}) back into the action (\ref{Sexpand}), expanding to second order in perturbations and dropping some boundary terms, we obtain the quadratic action for scalar perturbations 
\baq
\label{Squadr}
S^{(2)} &=& \frac{1}{2}\int d^4x\; a^3\left(\rule{0pt}{4ex}\right.
\left(\dot{\delta\phi}^2-\partial^{i}\delta\phi\partial_{i}\delta\phi+ 3H^2(2\epsilon_{\phi}-\eta_{\phi}) \delta\phi^2\right)\left(1+{\cal O}\left(\epsilon_{\phi},\h^2\right)\right)
\\\nonumber
&&+
\left(\dot{\delta{h}}^2-\partial^{i}\delta{h}\partial_{i}\delta{h}- (U''(h_0)+ 12\xi H^2)\delta{h}^2\right)\left(1+{\cal O}\left(\epsilon_{\phi},\h\sqrt{\epsilon_{\phi}}\right)\right)
\\\nonumber  
&&+ ~{\cal O}\left(\delta\phi\delta h H^2\sqrt{\epsilon_{{h}}\epsilon_{\phi}},\, \delta\phi\delta h H^2 \h \sqrt{\epsilon_{\phi}},\, \dot{\delta \phi} \delta h H \sqrt{\epsilon_{\phi}\epsilon_h}\right)
\left)\rule{0pt}{4ex}\right.~.
\eaq
Details of the computation are given in the Appendix \ref{fullaction}.

As can be seen in (\ref{Squadr}), any mixing between the Higgs and inflaton (or metric) fluctuations is suppressed by slow-roll and $\h$. To leading order in these small parameters the Higgs and inflaton fluctuations decouple: the inflaton part  
coincides with the single-field result \cite{Maldacena:2002vr} and the Higgs-dependent part reduces to the result (\ref{S_h}) obtained by treating the metric as a non-fluctuating background. This is the main result of our paper. For clarity, we recapitulate it explicitly in the next subsection stressing also frame-independency of the outcome. 

\subsection{Comparison to results on a fixed metric and inflaton background}
\label{sec:comp}
With the results of sections \ref{sec:JE} and \ref{sec:fluc} we can now perform a comparison of the difference
between including (and quantising) all fluctuations to quantising only the Higgs fluctuations in the Jordan frame action (\ref{S_intro}) on background with a non-fluctuating inflaton and metric. For this it is sufficient to compare the actions for the quadratic Higgs fluctuations. Denoting the Jordan frame action expanded to quadratic order in Higgs fluctuations $\delta h$ with the mean fields $g_0^{\mu\nu}$ and $\phi_0$ as $S^{(2)}_h\left(g_0^{\mu\nu},\phi_0,h_0+\delta h\right)_{\rm Jordan}$ from (\ref{S_intro}) we can write
\ee{S^{(2)}_h\left(g_0^{\mu\nu},\phi_0,h_0+\delta h\right)_{\rm Jordan}=-\f{1}{2}\int d^4x\; a^3\delta h\bigg(-\Box+ U''(h_0)+\xi R\bigg)\delta{h}\,.\label{eq:quad}}
Making use of equation (\ref{S_h}) one can then derive the result to quadratic order for the Einstein frame action (\ref{Sexpand}), which when again quantising only the Higgs fluctuations up to small terms coincides with the Jordan frame result
\ee{S^{(2)}_h\left(g_0^{\mu\nu},\phi_0,h_0+\delta h\right)_{\rm Einstein}=S^{(2)}_h\left(g_0^{\mu\nu},\phi_0,h_0+\delta h\right)_{\rm Jordan}+\mathcal{O}(\xi\alpha^2)\,.}
Finally, in the previous subsection we expanded the action to quadratic order in all fluctuations via the ADM formalism resulting in (\ref{Squadr}), from which we obtain our main result
\ea{S^{(2)}_h\left(g_0^{\mu\nu}+\delta g^{\mu\nu},\phi_0+\delta \phi,h_0+\delta h\right)_{\rm Einstein}&=S^{(2)}_h\left(g_0^{\mu\nu},\phi_0,h_0+\delta h\right)_{\rm Einstein}\nonumber \\ &=S^{(2)}_h\left(g_0^{\mu\nu},\phi_0,h_0+\delta h\right)_{\rm Jordan}\,,}
up to small terms of order
\ee{\bigg\{\sqrt{\epsilon_h\epsilon_\phi}\,,\xi\h\sqrt{\epsilon_\phi}
\,,\xi\h^2
\bigg\}
\lesssim \h\,.\label{eq:supp}}
We point out that at the level of the action when all fluctuations are quantised the choice of frame only amounts to a choice of a convenient set of variables and one should have
\ee{S^{(2)}_h\left(g_0^{\mu\nu}+\delta g^{\mu\nu},\phi_0+\delta \phi,h_0+\delta h\right)_{\rm Einstein}=S^{(2)}_h\left(g_0^{\mu\nu}+\delta g^{\mu\nu},\phi_0+\delta \phi,h_0+\delta h\right)_{\rm Jordan}\,.}
The size of the suppression $\sim \alpha$ depends on the energetic significance of the Higgs field $h_0$, but whenever it may be considered as a spectator the suppression is expected to be significant. Our result may be trivially generalised to all spectator scalar fields.
\section{Implications for the Standard Model Higgs}
Now we can analyse the specific case of the SM Higgs with a metastable vacuum. 

For a generic scalar field $\varphi$ with the action $S[\varphi,g^{\mu\nu}]$ we can write the generating functional on a curved background as
\ee{Z[J]=\int \mathcal{D}\varphi~e^{iS[\varphi,g^{\mu\nu}]+i\int d^4x\sqrt{-g}~J\varphi}.} The effective action denoted as $\Gamma[\langle\hat{\varphi}\rangle,g^{\mu\nu}]\equiv\Gamma[\varphi,g^{\mu\nu}]$, can be derived with a Legendre transformation \cite{{Peskin:1995ev}}
\ee{\Gamma[\varphi,g^{\mu\nu}]\equiv\int d^4x \sqrt{-g}~\mathcal{L}_{\rm eff}[\varphi,g^{\mu\nu}]\equiv-i\log Z[J]-\int d^4x\sqrt{-g}~J\varphi\equiv -U_{\rm eff}(\varphi)\int d^4x \sqrt{-g} \,,\label{eq:Leg}} where in the last step in order to obtain the effective potential we have assumed a constant field $\varphi$. 
The above definition can be generalised to all types of fields and for a result quadratic in fluctuations such as in (\ref{eq:quad}) by implementing the standard formulae for Gaussian path integrals \cite{Peskin:1995ev} one may write the 1-loop quantum contribution to the effective potential for the SM Higgs symbolically as \cite{vacstab7}
\ee{U^{\rm (2)}_{\rm eff,\, SM}(h_0)=-\f{i}{2}\sum_{i}n_i{\rm Tr\,\, log}\big[-\Box+M^2_i\big]\,,}
where the sum extends over all SM degrees of freedom that are counted by $n_i$ and the constant volume factor resulting from integration  visible in (\ref{eq:Leg}) is set to unity. If we now parametrize the Higgs doublet as 
\ee{H = \f{1}{\sqrt{2}}\left(\begin{array}{c} -i(\chi_1 - i \chi_2) \\ h_0+ (h + i \chi_3)\end{array}\right)\,,}
where $h_0$ is again the possible vacuum expectation value, the effective masses $M^2_i$ for the Higgs sector can be read from the definition of $U(h)$ from (\ref{higgspot})
\ee{M^2_h = -m^2 +3\lambda h_0^2 +\xi R\,\qquad M^2_{\chi_i} = -m^2 +\lambda h_0^2 +\xi R\,.}
Hence one may see that the non-minimal term for the Higgs $\sim \xi R H^\dagger H$ gives rise to an effective mass contribution.

The analysis of the section (\ref{sec:comp}) then shows that even if one were to quantize all fluctuations the resulting effective potential can nonetheless be approximated by quantising only the matter fields as defined in the Jordan frame action (\ref{S_intro}), where the error is given by the small terms in (\ref{eq:supp}).
The central values of the currently measured SM parameters are consistent with an instability at the scale $\Lambda_I\sim 10^{10}$GeV --  $10^{11}$GeV \cite{Degrassi:2012ry, Buttazzo:2013uya}, so we can from (\ref{eq:supp}) estimate the error coming from neglecting the metric and inflaton fluctuations to be smaller than
\ee{\alpha=\f{h_0}{M_{\rm pl}}\sim \f{\Lambda_I}{M_{\rm pl}}\lesssim 10^{-7}\,.}
This small number also describes the magnitude of a potential frame dependence of the result\footnote{To be precise, our argument does not imply the existence of a frame dependence but rather that even if it exists it must be subdominant for the problem at hand. }. 

To summarize, when investigating Higgs dynamics during inflation to a very good approximation one may treat gravity as a classical background given by some inflaton sector and the non-minimal term $\sim \xi R H^\dagger H$ as an effective mass contribution as defined by the Jordan frame action. This contradicts the claims in Ref. \cite{Calmet:2017hja}. The effect from the classical background curvature is however non-negligible and a flat space approximation for the effective potential misses relevant physics as demonstrated in \cite{vacstab7}. This topic we will further address in \cite{inprep}.
\section{Summary and conclusions}
The focus of this work has been to determine the validity of treating gravity as a non-fluctuating classical background when discussing the fluctuations of the Higgs field and the potential EW vacuum instability during high scale inflation. What we found was that the corrections from quantum gravity were negligible with a suppression of $10^{-7}$ or more and similarly that the quantization of the matter fields could equally well be made in the Einstein or Jordan frame.

We first explicitly showed that at the background level up to small terms the action in the Jordan frame coincides with the Einstein frame one. The result is contingent on the requirement of the Higgs field being a subdominant spectator and does not hold when it is a non-negligible component in the energy-density for example as in Higgs inflation. We then proceeded to perform a full ADM fluctuation expansion on the action to quadratic order. Similarly to the background approach the result conclusively showed that up to small terms the result coincided with what one would obtain by neglecting the quantum nature of the metric.

In hindsight our results were entirely to be expected: the best fit value for the instability occurs at a scale that is seven orders of magnitude below $M_{\rm pl}$, which is precisely the suppression given by our calculation. However, recently in \cite{Calmet:2017hja} it was claimed that a Jordan frame quantization of only the matter fields as was implemented in \cite{vacstab7} and \cite{Herranen:2015ima} is incomplete. The analysis of this work shows this statement to be incorrect. We conclude that quantum field theory on a classical background is a well-suited approach for analysing the various issues related to the early Universe vacuum instability.

\acknowledgments{The research leading to these results has received funding from the European Research Council under the European Union's Horizon 2020 program (ERC Grant Agreement no.  648680). AR was supported by STFC grant ST/L00044X/1. AR and TM were  supported by STFC grant ST/P000762/1}

\appendix
\section{The quadratic action for scalar perturbations}
\label{fullaction}

We expand the action (\ref{Sexpand}) with the ADM decomposition of the metric and choosing the gauge (\ref{gauge}). Tensor perturbations do not couple to scalars at this order according to the standard scalar-vector-tensor decomposition of linear perturbations \cite{Mukhanov:1990me} and a straightforward computation yields the quadratic action for scalar perturbations,  
\beq
\label{S2split}
S^{(2)} = S^{(2)}_R+S^{(2)}_K+ S^{(2)}_V~,
\eeq
where the individual terms are given by 
\baq
\label{S2R}
S^{(2)}_R &= & \int d^4x\; a^3\frac{1}{2}M_{\rm pl}^2 (N-1)
\left[\rule{0pt}{4ex}\right.
-6H^2(N-1)-4H\partial_i N^{i}
\left]\rule{0pt}{4ex}\right.~,\\
\label{S2K}
S^{(2)}_K &= & \int d^4x\; a^3
\left[\rule{0pt}{4ex}\right.
\frac{1}{2}\left(\dot{\delta\phi}^2+\dot{\delta h}^2-\partial^{i}\delta\phi\partial_{i}\delta\phi-\partial^{i}\delta h\partial_{i}\delta h\right)
-N^{i}\left(\dot{\phi}\partial_i\delta\phi+\dot{h}\partial_i\delta h\right)\\\nonumber
&&
+\frac{1}{2}(N-1)^2\left(\dot{\phi}^2+\dot{h}^2\right)-(N-1)\left(\dot{\phi}\dot{\delta\phi}+\dot{h}\dot{\delta h}\right)
\left]\rule{0pt}{4ex}\right.~,\\
\label{S2V}
S^{(2)}_V &= & \int d^4x\; a^3
\left[\rule{0pt}{4ex}\right.
-\frac{1}{2}V''(\phi)\delta\phi^2 -\frac{1}{2}\left(U''(h)+\xi R\right)\delta h^2-\frac{1}{2}\xi\h^2 \partial^{i}\delta\phi\partial_{i}\delta\phi\\\nonumber
&&-\xi\h\frac{\delta h}{M_{\rm pl}}\left(-2\dot{\phi}\dot{\delta\phi}+4V'(\phi)\delta\phi\right)-\frac{1}{2}\xi\h^2\left(-\dot{\delta\phi}^2+2\dot{\phi}N^{i}\partial_{i}\delta\phi+2V''(\phi)\delta\phi^2\right)\\\nonumber
&&
+(N-1)\left(-V'(\phi)\delta\phi-U'(h)\delta h-\xi\h\frac{\delta h}{M_{\rm pl}}\left(\dot{\phi}^2+4V(\phi)\right)-\xi\h^2\left(\dot{\phi}\dot{\delta\phi}+2V'(\phi)\delta\phi\right)\right)\\\nonumber
&&+\frac{1}{2}\xi\h^2(N-1)^2\dot{\phi}^2
\left]\rule{0pt}{4ex}\right.~.
\eaq
where in the above $R$ is given by the background metric. In arriving at eq. (\ref{S2R}) we have performed partial integrations and used that the shift $N^i$ can be expressed in the form $N^i = \partial^i \psi$, where $\psi$ is a first order scalar perturbation.

The lapse $N$ and shift $N^i$ act as Lagrange multipliers; their equations of motion $\delta S^{(2)}/\delta N = 0$ and $\delta S^{(2)}/\delta N^i = 0$ contain no time derivatives. To first order in perturbations the solution of these constraint equations is given by 
\baq
\label{constraints_full}
N &=& 1 +\frac{\dot{\phi}(1+\xi\h^2)}{2 H M_{\rm pl}^2}\delta\phi+\frac{\dot{h}}{2 H M_{\rm pl}^2}\delta h~,\qquad N^i = \partial^i \psi~, \\\nonumber 
\partial^i\partial_i \psi &=& -\frac{\dot{\phi}}{2 H M_{\rm pl}^2}\dot{\delta\phi}(1+\xi\h^2)-\frac{\dot{h}}{2 H M_{\rm pl}^2}\dot{\delta h} \\\nonumber
&&
+\frac{\delta\phi}{2HM_{\rm pl}^2}\left(\frac{\dot{\phi}(1+\xi\h^2)}{2 H M_{\rm pl}^2}\left(-6H^2M_{\rm pl}^2 +(1+\xi\h^2)\dot{\phi}^2+\dot{h}^2\right)-V'(\phi)(1+2\xi\h^2)\right)\\\nonumber
&&
+\frac{\delta h}{2HM_{\rm pl}^2}\left(\frac{\dot{h}}{2 H M_{\rm pl}^2}\left(-6H^2M_{\rm pl}^2 +(1+\xi\h^2)\dot{\phi}^2+\dot{h}^2\right)-U'(h)-\xi\h\frac{4V(\phi)+\dot{\phi}^2}{M_{\rm pl}}\right)\,.
\eaq
No slow-roll assumptions have been made up to this point.

By substituting the solutions for $N$ and $N^i$ back into eqs. (\ref{S2R}) - (\ref{S2V}), we get the full quadratic action for scalar perturbations (note that parts proportional to $\delta h\dot{\delta h}$ and $\delta \phi\dot{\delta \phi}$ can be brought to form $\delta h^2$ and $\delta\phi^2$ by partial integration). 

For our purposes it is however not necessary to do this step explicitly. Our main goal is to show that terms generating Higgs-inflaton mixing $\delta\phi\delta h$ are suppressed compared to the $\delta h^2$ part of the quadratic action in the limit where the Higgs is a subdominant spectator. To this end we only need to keep track on the magnitude of the mixing terms. From eqs. (\ref{S2R}) - (\ref{S2V}) and (\ref{constraints_full}) we see that the mixing terms (to leading order in  $\h$) are proportional to  
\baq
\label{mixing}
S^{(2)}&\supset& 
\left\{\rule{0pt}{4ex}\right.
\int d^4x\; a^3 \dot{\delta\phi}\delta h\;
 {\cal O}\left(\frac{\dot{\phi}\dot{h}}{HM_{\rm pl}^2}\right), ~\int d^4x\; a^3 \dot{\delta h}\delta \phi\;
 {\cal O}\left(\frac{\dot{\phi}\dot{h}}{HM_{\rm pl}^2}\right), \\\nonumber
&&\int d^4x\; a^3 \delta\phi\delta h\;
 {\cal O}\left(\frac{\dot{\phi}\dot{h}}{M_{\rm pl}^2},\frac{\dot{\phi}^3\dot{h}}{H^2M_{\rm pl}^4},\frac{\dot{\phi}\dot{h}^3}{H^2M_{\rm pl}^4},\frac{\dot{\phi} U'(h)}{HM_{\rm pl}^2},\frac{\dot{h} V'(\phi)}{HM_{\rm pl}^2}, \frac{\h \dot{\phi} V(\phi)}{H M_{\rm pl}^3}, 
 \frac{\h \dot{\phi}^3}{H M_{\rm pl}^3}\right)
\left\}\rule{0pt}{4ex}\right.~.
\eaq
We recap that no slow-roll has been assumed to arrive at this result. 

\subsection{Inflaton dominated slow-roll limit}

We now concentrate on inflaton dominated slow-roll solutions where the Higgs is an energetically subdominant light spectator. From the background solutions (\ref{Heq}), ({\ref{phieq}) and (\ref{chieq}), and the definition of the slow-roll parameters (\ref{srparameters}) it follows that
\beq
\label{App:SR_relations}
\frac{V'(\phi)}{3H^2 M_{\rm pl}} = \sqrt{2\epsilon_{\phi}}\left(1+{\cal O}(\h^2)\right)~,\qquad \frac{U'(h)}{3H^2 M_{\rm pl}} = \sqrt{2\epsilon_h} -4\xi\h\left(1+{\cal O}(\epsilon,\rho_{h}/\rho_{\rm tot})\right)~. 
\eeq

Using these together with eqs. (\ref{Heq}), ({\ref{phieq}) and (\ref{chieq}) in eq. (\ref{constraints_full}), the solutions for the lapse and shift become 
\baq
\label{constraints_SR}
N &=& 1 -\sqrt{\frac{\epsilon_{\phi}}{2}}\frac{\delta\phi}{M_{\rm pl}}\left(1+{\cal O}(\h^2)\right)+{\cal O}(\sqrt{\epsilon_h})\frac{\delta h}{M_{\rm pl}}~,\qquad N^i = \partial^i \psi~,\\\nonumber 
\partial^i\partial_i \psi &=& \sqrt{\frac{\epsilon_{\phi}}{2}}\frac{\delta\dot{\phi}}{M_{\rm pl}}\left(1+{\cal O}(\h^2)\right)+{\cal O}\left(\sqrt{\epsilon_h}\frac{\dot{\delta h}}{M_{\rm pl}}, \epsilon_{\phi}^{3/2}\frac{H\delta \phi}{M_{\rm pl}}, 
\epsilon_{\phi}\sqrt{\epsilon_h}\frac{H\delta h}{M_{\rm pl}}, \h\frac{H\delta h}{M_{\rm pl}}\right)~.
\eaq
Noting that $\sqrt{\epsilon_{h}} = {\cal O}(\alpha \eta_h)$ and dropping the slow-roll and $\h$ suppressed terms we get the result (\ref{constraints}) quoted in the text.

The results for $\delta\phi^2$ and $\delta h^2$ terms in eq. (\ref{Squadr}) follow directly by substituting the constraints (\ref{constraints_SR}) in eqs. (\ref{S2R}) - (\ref{S2V}) and using the 
backgrounds solutions (\ref{Heq}), ({\ref{phieq}) and (\ref{chieq}) together with eq. (\ref{App:SR_relations}). 

Finally, using eqs.  (\ref{Heq}), ({\ref{phieq}), (\ref{chieq}) and (\ref{App:SR_relations}) in eq. (\ref{mixing}) we find that the leading mixing terms scale proportional to 
\baq
S^{(2)}&\supset& 
\left\{\rule{0pt}{4ex}\right.
\int d^4x\; a^3 H\dot{\delta\phi}\delta h\;
 {\cal O}\left(\sqrt{\epsilon_{\phi}\epsilon_{h}}\right), ~\int d^4x\; a^3 H\dot{\delta h}\delta \phi\;
 {\cal O}\left(\sqrt{\epsilon_{\phi}\epsilon_{h}}\right), \\\nonumber
&&\int d^4x\; a^3 H^2\delta\phi\delta h \;
 {\cal O}\left(\sqrt{\epsilon_{\phi}\epsilon_{h}},\h\sqrt{\epsilon_{\phi}}\right)
\left\}\rule{0pt}{4ex}\right.~,
\eaq
as written in eq. (\ref{Squadr}).

\end{document}